# High performance current and spin diode of atomic carbon chain between transversely symmetric ribbon electrodes

Yao-Jun Dong[1], Xue-Feng Wang[1,2,*], Shuo-Wang Yang[3], Xue-Mei Wu[1,2]

[1] College of Physics, Optoelectronics and Energy, Soochow University, Suzhou 215006, China

[2] State Key Laboratory of Functional Materials for Informatics and Key Laboratory of Terahertz Solid-State Technology, Shanghai Institute of Microsystem and Information Technology, Chinese Academy of Sciences, 865 Changning Road, Shanghai 200050, China

[3] Institute of High Performance Computing, A*Star, 1 Fusionopolis Way, 16-16 Connexis, Singapore 138632, Singapore

*E-mail: wxf@suda.edu.cn

## Abstract

We demonstrate that giant current and high spin rectification ratios can be achieved in atomic carbon chain devices connected between two symmetric ferromagnetic zigzag-graphene-nanoribbon electrodes. The spin dependent transport simulation is carried out by density functional theory combined with the non-equilibrium Green's function method. It is found that the transverse symmetries of the electronic wave functions in the nanoribbons and the carbon chain are critical to the spin transport modes. In the parallel magnetization configuration of two electrodes, pure spin current is observed in both linear and nonlinear regions. However, in the antiparallel configuration, the spin-up (down) current is prohibited under the positive (negative) voltage bias, which results in a spin rectification ratio of order $10^4$. When edge carbon atoms are substituted with boron atoms to suppress the edge magnetization in one of the electrodes, we obtain a diode with current rectification ratio over $10^6$.

**Introduction**

The current rectification in a diode is fundamental in electronics and the spin current rectification is equally important in spintronics. In the diffusive transport regime, various charge and spin diodes have been proposed[1-4]. Since the pioneer work done by Aviram and Ratner on molecular rectifier[5], rectification systems in atomistic scale have been extensively studied due to the scientific and technologic interests in quantum transport and molecular electronics[6,7]. Ballistic spin diodes have also been proposed, including two quantum wire electrodes coupled by a tunnel barrier in antiparallel magnetic configuration[8] and molecular wires bridging magnetic and nonmagnetic electrodes[9]. Recently, large current rectification ratio has been detected in longitudinally asymmetric but transversely symmetric diblock dipyrimidinyldiphenyl molecules connected[10, 11]. Meanwhile, a device composed of organic magnetic molecules sandwiched by gold electrodes[10] has been expected to obtain a spin rectification ratio over 100. A half-metallic behavior in one dimensional infinite chromium porphyrin array has been predicted[11].

Carbon based electronics and spintronics have been attracting a huge amount of interest especially after the discovery of graphene[12-19]. Graphene has advantageous electronic properties and can be easily patterned into designed nanostructures and nanodevices such as nanoribbons and field effect transistors[19-24]. On the other hand, atomically perfect narrow graphene nanoribbons (GNRs) can be fabricated by bottom-up methods like self-assembly[25,26]. In addition, the spin current can be injected into graphene from ferromagnetic electrodes at room temperature[27,28]. Particularly, the zigzag graphene nanoribbons (ZGNRs) have been the focus of many studies due to its tunable band gap and intrinsic edge magnetism[29,30], and giant magneto-resistance devices of graphene nanoribbon were predicted and fabricated[31,32]. Various methods were proposed to enhance the charge and spin rectification as well as the spin polarization in ZGNR devices[33-36]. A current rectification ratio up to 6000 is predicted in diarylethene molecule sandwiched between a ZGNR and an armchair graphene nanoribbon[37]. The spin rectification ratio up to 100 in bent graphene nanoribbons has been predicted based on ab initio simulation[38].

Recently, carbon atomic chains (CACs) have been successfully fabricated from graphene with the help of a high energy electron beam[39-41] and its transport properties were also characterized experimentally[42]. The advancement of such technology is significant since CACs may be used as transport channels or on-chip interconnects for graphene-based devices, nanoelectronic or spintronic nanodevices[29,43-45]. *Ab initio* simulations have indicated that the CACs between ZGNRs exhibit similar electron transport properties as those between metal electrodes[46-48] and can be employed as spin filters or spin valves[49,50]. It is predicted that in C doped BN atomic wire connected between gold electrodes via S atoms[51], a current rectification ratio can be enhanced up to 130. In the current work, we elucidate the importance of geometry symmetry match among CACs and the two electrodes in enhancing the rectification ratio in atomistic systems. We demonstrate how to realize a high performance spin filter with high spin rectification ratios as well as giant current rectification ratio in transversely symmetric ZGNR-CAC systems.

**Computational Models and Methods**

We consider device systems where a CAC is connected between two transversely symmetric electrodes of pristine or doped 6-ZGNRs shown in Figures 1(a), 3(a), 4(a) and 5(a). A 7-atom CAC is employed in the simulation since ballistic transport maintains in the chains with an odd number of atoms[52] and we have verified that our results/conclusions are not sensitive to the length of the chain. Three-membered carbonic rings are designed to link between the CACs and the ZGNRs to keep the geometry symmetry of the system. The two-probe device is partitioned into the left (L) and right (R) electrodes and the central scattering region where a buffer layer of 6-ZGNR is put at each ends of the CAC to ensure a negligible contact effect on the electrostatic potential in the electrodes. The edges of the electrode ZGNRs are assigned in ferromagnetic states[27,31] and the left electrode is in either antiparallel (FMa) or parallel (FMp) magnetization configurations to the right electrode.

In the doped ZGNRs, the edge C atoms are replaced by dopant atoms of element B or N. The dangling bonds of the edge atoms are terminated by H atoms. In Table 1 we present the formation energies per primitive cell $E_{form} = E_{nH\text{-}ZGNR} - (E_{ZGNR} + nE_{H2}/2)$ of pristine, B-doped, and N-doped 6-ZGNRs with each edge atom terminated by $n$ H atoms, $n = 1$ or 2. Here $E_{nH\text{-}ZGNR}$ is the total energy per primitive cell of ZGNR terminated by $n$ H atoms, $E_{ZGNR}$ is the total energy per primitive cell of ZGNR without H atom, and $E_{H2}$ is the total energy of a $H_2$ molecule. Pristine ZGNRs prefer 2H termination ($sp^3$) while doped ZGNRs prefer 1H termination ($sp^2$). Since 1H termination is also feasible in pristine ZGNRs by controlling the chemical potential of hydrogen via temperature and pressure of $H_2$ gas[53,54] or by the hydrogen plasma etching method[55], here we assume 1H termination in all the cases.

Table 1: Formation energies per primitive cell of ZGNRs in which each edge atom is terminated by one H atom or two H atoms.

| Element of Edge Atoms (X) | C | B | N |
|---|---|---|---|
| 1H-X (eV) | -6.528 | -4.165 | -1.557 |
| 2H-X (eV) | -6.936 | -3.847 | 1.274 |

The geometry optimization for each system is performed firstly by the Vienna *ab initio* simulation package (VASP)[56, 57] and then by the Atomistix Toolkits (ATK)[59, 60] until the forces on each atom is less than 0.02 eV/Å. The structure parameters are given in Fig. S1. Secondly, the electron transports are calculated by the nonequilibrium Green's functions method (NEGF) combined with *ab initio* density functional theory (DFT) implemented in the ATK package[58, 59]. We use the Perdew-Zunger parametrization of the local spin density approximation (LSDA) for the exchange correlation functionals, which leads to the same conclusion as the spin dependent generalized gradient approximation with the Perdew-Burke-Ernzerhof parametrization (SGGA-PBE). The wave functions are expanded in a basis set of double-ζ orbitals plus one polarization orbital (DZP), which can preserve an accurate description of π and π* bonds. A 15Å thick vacuum layer is adopted to separate the ribbons in neighbouring supercells which is enough to suppress the of the coupling between them. In the transport calculation, the energy cutoff is 2000 eV, the mesh grid in the *k*-space is of size 1×1×100, and an electronic temperature of 300K is used in the technique of the real-axis integration for the nonequilibrium Green's functions.

The current ($I_\sigma$) of spin $\sigma$ through the device is evaluated by the Landauer-Büttiker formula: $I_\sigma = \frac{e}{h} \int_{-\infty}^{\infty} T_\sigma [f(E-\mu_R) - f(E-\mu_L)] dE$. Here $T_\sigma(E)$ is the transmission spectrum and $f(E-\mu_X)$ is the Fermi distribution function with the Fermi energy $\mu_X$ in electrode X (L or R). Under a voltage bias $V_b$, the current flows from the left to the right electrode and we assume $\mu_L = -eV_b/2$ and $\mu_R = eV_b/2$. The integration is over the transport window in energy $E \in [-e|V_b|/2, e|V_b|/2]$ at zero temperature and mainly the transmission spectra in the transport window contribute to the current even at finite temperature. The linear conductance of the device reads $G_\sigma = \frac{e^2}{h} T_\sigma$.

**Results and discussion**

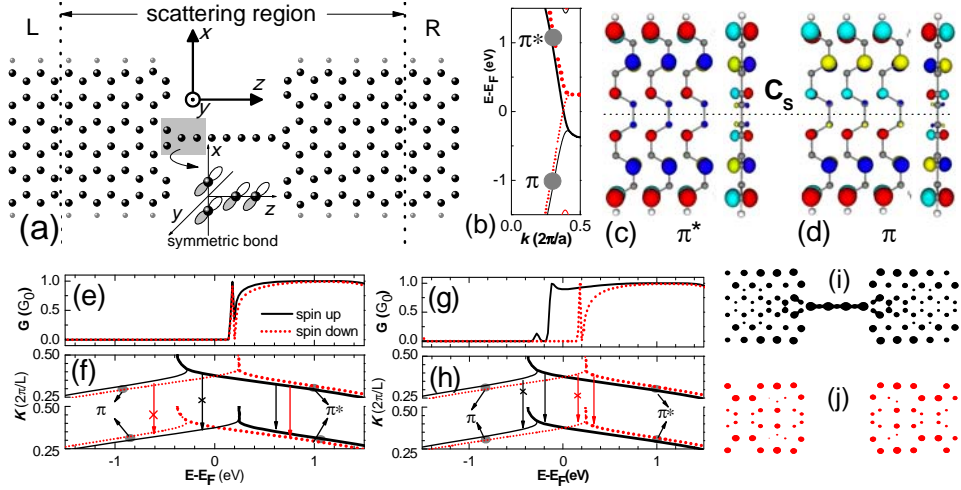

Figure 1 **(a)** Schematic illustration of the pristine device system and its partitions in the calculation. The big dark (small light) spheres indicate the C (H) atoms. The left electrode is assigned magnetized-up. For a transport channel, the electronic orbitals of the four C atoms in the contact area (shadow) have a symmetric form as shown in the inset. **(b)** The energy bands of pristine 6-ZGNRs. The black solid (red dotted) curves are for spin up (down) bands and the thick (thin) curves for π* (π) states. **(c)** and **(d)** the electronic wave function isosurfaces of the spin-up (π*) and the spin-down (π) bands near the Fermi energy, respectively. **(e)** The linear spin-up and spin-down conductance spectra in the FMa configuration. **(f)** The available (arrows) and prohibited (crossed arrows) electron transport channel from states in electrodes L to those in electrode R. **(g)** and **(h)** the linear spin-up and spin-down conductance spectra in FMp configuration. **(i)** and **(j)** The spin-up and spin-down local density of states respectively, at the Fermi energy in the FMp configuration. The three bond lengths from the end to the center of the seven-atom CAC are 1.354, 1.290, and 1.312 Å, respectively. The bond lengths in the contact between CAC and GNRs electrodes are all 1.447 Å.

Our calculation results show that the pristine 6-ZGNRs are metallic at the ferromagnetic state and have an energy band structure near the Fermi energy as shown in Figure 1(b) when being assumed up-magnetized. The corresponding states are composed of $p_y$ orbitals and are antisymmetric with respect to the *x-z* plane. In addition, the wave functions of the π* bands are antisymmetric and those of the π band symmetric with respect to the *y-z* plane as shown in Figure 1(c) and (d), respectively. The spin up/down π* band crosses with the spin up/down π band near 0.2eV above/below the Fermi energy at $k = 0.4 \times 2\pi/a$ with *a* = 2.461Å the lattice constant of the ZGNR. The two bands then twist with each other and become almost degenerate at higher k when the corresponding wave functions are confined to the edges of the ZGNR. Interestingly, the two spin up bands (black solid) decrease in energy at $k > 0.4 \times 2\pi/a$ while the spin down ones (red dotted) become flat.

In CACs where the C atoms are bonded to each other via the *sp* hybridization, the wave functions of states near the Fermi energy are composed of carbon $p_x$ or $p_y$ orbitals. The $p_x$ ($p_y$) orbitals are symmetric with the *x* (*y*) axis and antisymmetric with respect to the plane perpendicular to the *x* (*y*) axis. When a CAC is connected to a symmetric ZGNR electrode along its central line as shown in Figure 1(a), those electrons passing through the device are required to have a wave function symmetric with respect to the *y-z* plane and antisymmetric with respect to the *x-z* plane as indicated in the inset.

The conductance spectra in the linear response region are presented in Figure 1(e) and (g) for the FMa and FMp configurations of the electrode magnetizations, respectively. In the FMa configuration, for each spin there is an energy difference about 0.4eV between the cross points of π and π* bands in the two electrodes as shown in Figure 1(e) and (f). Because the electrons in the π band cannot pass through the CAC due to the transverse symmetry mismatch, the spin-up and spin-down conductance spectra show similar step forms. The step edges are located at 0.2eV above the Fermi energy, meanwhile, a sharp conductance peak appears next to the edges corresponding to the flat part of the energy bands. Note that a conductance dip follows the conductance peak and results in a conductance oscillation near the edges. This is an interference result of the transverse confinement mismatch between the wide ZGNRs and the narrow CAC. In the FMp configuration, the energy bands of the two electrodes are aligned with each other as shown in Figure 1(g) and (h). The spin-down conductance spectrum appears the same as that in the FMa configuration, while the step edge of the spin-up one shifts down to −0.2eV below the Fermi energy and the sharp peak next to the edge shifts to lower energy but the range becomes wider, resulting in a perfect spin transport window about 0.4eV around the Fermi energy. It means the system can work as a spin filter. Based on spin up and spin down local density of states (LDOS) at the Fermi energy presented in Figure 1(i) and (j), it is seen that the spin up electrons in the edge states of the electrodes can flow through the CAC. Meanwhile, the spin-down LDOS in the CAC at the Fermi energy is almost zero and there is no transport channel for the spin-down electrons to pass through the device.

Furthermore, we study the electronic conductance in this CAC system. The calculated spin-up (black) and spin-down (red) I-V curves in the FMa configuration are presented in Figure 2(a). In the linear regime, the device is OFF for both spin currents due to the symmetry mismatch between the electrodes as illustrated in Figure 1(f). Under a bias $V_b > 0$, the band structure of the left (right) electrode shift to higher (lower) energies. The step edge of the spin-up conductance spectrum shifts to higher energies accordingly and always remains outside the transport window as shown in Figure 2(c). Therefore, the spin-up transport channel is OFF. On the other hand, the step edge of the spin-down conductance spectrum shift to lower energy and moves into the transport window at $V_b = 0.2V$ [see Figure 2(d)]. The spin-down transport channel is then ON and the current increases quickly with $V_b$ until saturates at $V_b = 1.0V$. On the contrary, the spin-up transport channel becomes ON when $V_b < −0.2V$ and the current saturates at $V_b = −1.0V$ while spin-down transport channel remains OFF.

As discussed above, the device works as a high performance spin diode as well as a perfect spin filter. When the applied bias reverses, both the direction and the magnitude of the spin current remain unchanged. According to the definition of spin dependent rectification ratio, $RR_\sigma = [I_\sigma(V) - I_\sigma(-V)] / |\text{Min}\{I_\sigma(V), I_\sigma(-V)\}|$, we plot the rectification ratio versus the bias amplitude for each spin in Figure 2(b). The rectification ratio becomes higher than $10^4$ for $V_b > 0.2V$. The rectifying behavior in the ZGNR-CAC structures is similar to that in ZGNRs[32,34] but with much higher rectification ratio.

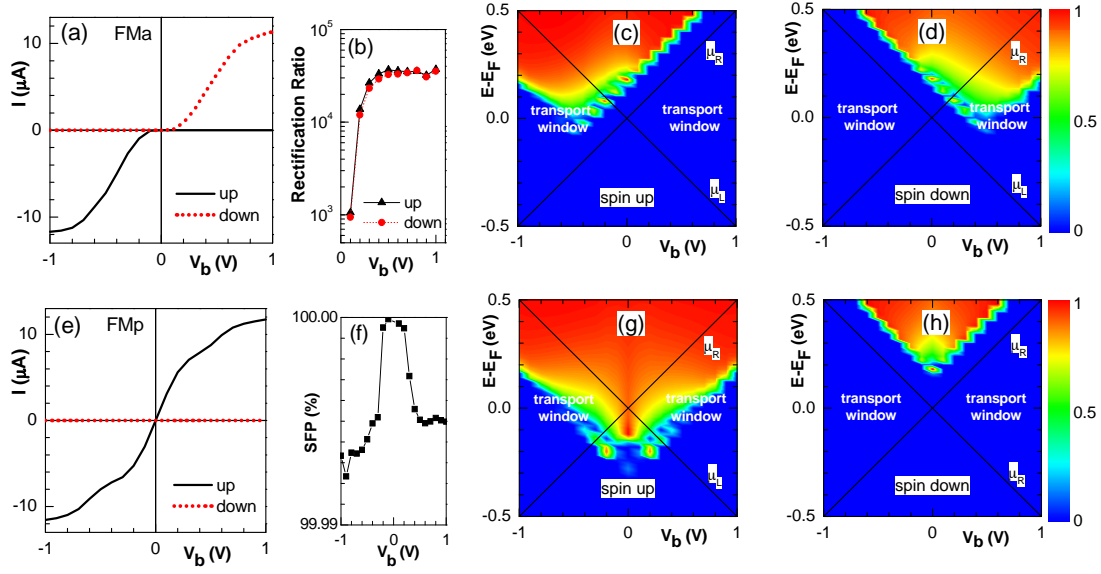

Figure 2 **(a)** I-V curves of spin-up (solid) and down (dotted) components for a CAC device with pristine 6-ZGNR electrodes in the FMa configuration. **(b)** The corresponding spin-up (triangles) and down (dots) rectification ratios versus $V_b$. **(c)** and **(d)** 3D plot of the spin-up and down transmission spectra on the E-$V_b$ plane, respectively. **(e)** The same as (a) in the FMp configuration. **(f)** The spin filtering polarization versus $V_b$ in the FMp configuration. **(g)** and **(h)** The same as (c) and (d), respectively, in the FMp configuration.

In addition, we also plot the *I-V* curve for the FMp configuration in Figure 2(e). For spin-up electrons, the device is metallic in the linear transport regime and the spin-up current increases monotonically with the bias until it saturates at $V_b = 1.0$V. Meanwhile, spin-down current keeps OFF. The spin filter polarization (SFP=$[(I_{up}-I_{down})/(I_{up}+I_{down})]\times 100\%$) approaches to 100% in the studied bias region [-1, 1] V as shown in Figure 2(f). In Figures 2(g) and (h) for spin-up and spin-down, respectively, we present the corresponding conductance spectrum under bias to illustrate the physical mechanism behind. Part of the spin-up $\pi^*$ bands in both electrodes is inside the transport windows and the spin-up transport is ON. In contrast, the spin-down $\pi^*$ band in electrode R shifts down with the increase of bias and is always localized outside of the transport windows. The spin-down channel is then OFF.

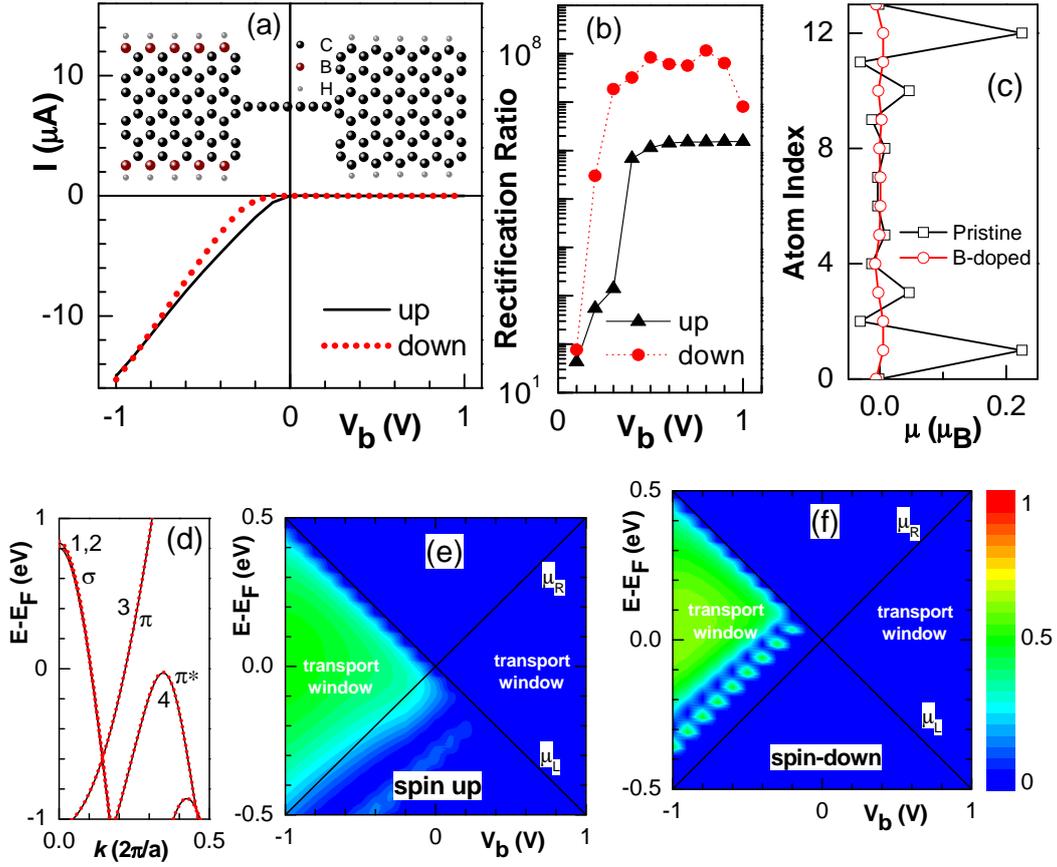

Figure 3 **(a)** I-V curves of spin-up (solid) and down (dotted) components in a device with left electrode doped by B atoms on both edges as schemed in the inset. The big black, big brown, and small grey spheres represent the C, B, and H atoms, respectively. **(b)** Spin-up and down rectification ratios versus $V_b$. **(c)** Atomic magnetic moment $\mu$ in B-doped (dots) and pristine (squares) 6-ZGNRs for atoms indexed from lower (index 0) to higher (index 13) edge. **(d)** Energy bands of B-doped 6-ZGNRs near the Fermi energy. **(e)** and **(f)** The spin-up and down transmission spectra, respectively.

Previous experiments have suggested that substitutional doping on the edges of graphene nanoribbons may greatly enhance the current rectification ratio and a ratio up to 16 has been measured in armchair nanoribbon[24]. Here we show that a giant current rectification ratio can be realized in a CAC device with edge C atoms in the left electrode substituted by B atoms as schemed in the inset of Figure 3(a). Its simulated I-V curves for both up and down spins in the FMp configuration are plotted in Figure 3(a) and the results are similar to that in the FMa configuration due to the magnetism suppression from B doped electrode. The device is ON within linear I-V curves under negative bias and the current is spin polarized at low bias. In contrast, the device is OFF under positive bias where the current becomes very small and decreases with the bias. As a result, the rectification ratio increases quickly with the bias as shown in Figure 3(b). The spin-up and spin-down rectification ratios can reach over $10^8$ and $10^6$ respectively; while, the average current rectification ratio is over $10^6$ under bias $|V_b| > 0.4$V.

In Figure 3(c), we plot the transverse distribution of the atomic magnetic moment across pristine (squares) and B-doped (dots) 6-ZGNRs. The doping suppresses the edge magnetic moment[55], and the 4 energy bands near the Fermi level in B-doped 6-ZGNRs become almost spin degenerate as plotted in Figure 3(d). Bands 1 and 2 are degenerate except at small $k$ and their wave functions are symmetric with respect to the $x$-$z$ plane, showing σ bonding characteristics (see Figure S2(a)-(b)). Bands 3 and 4 originate from $p_y$ orbitals but the wave functions in band 3 are antisymmetric with respect to the $y$-$z$ plane (see Figure S2(c)-(d)). So electrons near the Fermi energy in bands 1, 2, and 3 cannot reach to the right electrode of pristine 6-ZGNR due to the symmetry mismatch of wave functions. Because the top of Band 4 is slightly (~23meV) below the Fermi energy and locates ouside of the transport window $[-eV_b/2, eV_b/2]$ in the linear regime, the device is OFF when $V_b \geq 0$ as shown in Figure 3(a). When $V_b < 0$, Band 4 shifts to higher energies with the Fermi energy in the left electrode and overlaps with the spin-up π* band in the right electrode. The spin-up transmission spectrum fills almost uniformly the transport windows of negative bias as shown in Figure 3(f). The spin-up current shows a very small threshold voltage and then increases linearly with the bias. On the other hand, the spin-down π* band in the right electrode enters the transport windows only when bias $V_b < -0.2$V. The spin-down transmission spectrum has a shift to higher energy and negative bias with a bigger maximum compared to its spin-up counterpart. The spin-down current has a bigger threshold voltage and become almost the same at $V_b = -1$V.

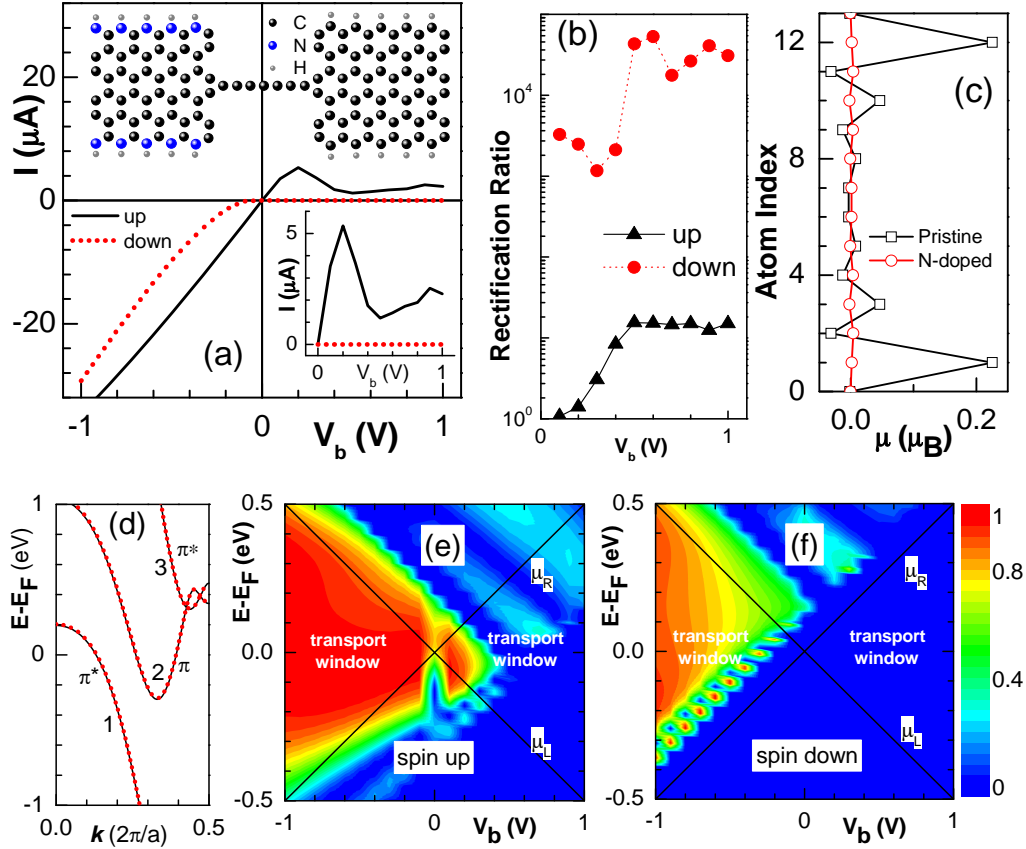

Figure 4 **(a)** I-V curves of spin-up (solid) and down (dotted) components in a device with left electrode doped by N atoms on both edges as schemed in the inset. The big black, big blue, and small grey spheres represent the C, N, and H atoms, respectively. **(b)** Spin-up and down rectification ratios versus $V_b$. **(c)** Atomic magnetic moment $\mu$ in N-doped (dots) and pristine (squares) 6-ZGNRs for atoms indexed from lower (index 0) to higher (index 13) edge. **(d)** Energy bands of N-doped 6-ZGNRs near the Fermi energy. **(e)** and **(f)** The spin-up and down transmission spectra, respectively.

Besides, we also study the N edge-doped effect and the results are given in Figure 4. As shown in Figure 4(a), the device is OFF for spin-down electrons when $V_b > 0$ and the negative differential resistance appears for spin-up electrons for $V_b \in [0.2, 0.5]$ V with a peak to valley ratio of 4.54. The resulting SFP is almost 100% and the device can work as a perfect spin filter effect when $V_b > 0$. While, when $V_b < 0$, the spin-down current shows semiconducting with a threshold voltage around 0.1 V but the spin-up one appears metallic. At this moment, it exhibits unique half-metallic. Both spins show linear *I-V* dependence at $V_b < 0$. Due to the block of spin-down current at $V_b > 0$, the device exhibit a giant rectification effect with a ratio around $10^4$ for spin down electrons as shown in Figure 4(b). The current rectification ratio vanishes at $V_b=0$ and may reach 10 at $V_b=1$ V. Similar to that in the B doped case, the atomic magnetic moment μ in the N edge-doped 6-ZGNR is also suppressed as shown in Figure 4(c) and the left electrode appears nonmagnetic with spin degenerate energy bands as illustrated in Figure 4(d). There are three bands (1, 2 and 3) near the Fermi energy and all of them are composed of $p_z$ orbitals. Bands 1 and 3 show π* characteristics and their wavefunctions are symmetric with respect to the y-z plane (see Figure S2(e) and (g)). Band 2 is a result of π bonding and doesn't contribute to the electronic transport (see Figure S2(f)). Because only electrons in π* can pass through the CAC due to the symmetry selection rule, the spin-up conductance spectrum at $V_b=0$ originates from Band 1 near the Fermi energy as shown in Figure 4(e), and the spin-down conductance spectrum from band 3 as shown in Figure 4(f). With the variation of $V_b$, the energy bands of the two electrodes shift against each other in energy and the conductance spectra change accordingly as illustrated in Figure 4(e) and (f).

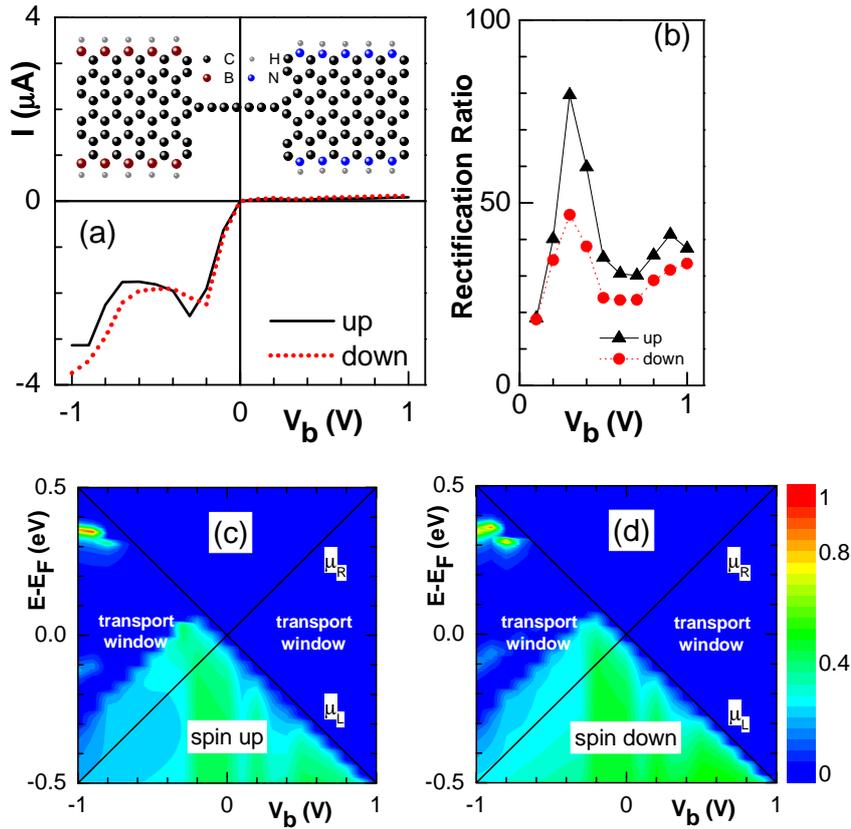

Figure 5 **(a)** I-V curves of spin-up (solid) and down (dotted) components in a CAC device with left B-doped and right B-doped electrodes as schemed in the inset **(b)** Spin-up and down rectification ratios versus $V_b$. **(c)** and **(d)** The spin-up and down

Finally, we study a P/N junction composed of B doped left electrode and N doped right electrodes as schemed in the inset in Figure 5(a) together with calculated I-V curves. The device is OFF under positive bias with a current about 0.1 μA at $V_b$ = 1.0 V. However, this value is much larger that ( ~ $10^{-6}$ μA at $V_b$ = 1.0V) in Figure 3(a) for the system with only one electrode doped by B atoms. As a result, the rectification ratio as shown in Figure 5(b) is much lower with the maximum spin-up (down) rectification ratio 80 (47) at 0.3V. Interestingly, the top of Band 4 in the left electrode (Figure 3(d)) matches in energy with the top of band 3 in the right electrode (Figure 4(d)) at $V_b$ =−0.25 V. The current reaches a maximum and then a NDR effect occurs in the bias range [−0.5, −0.25] V.

**Summary**

Based on first-principle simulations, we have shown that transversely symmetric ZGNR-CAC-ZGNR structures are promising materials for high-performance charge and spin rectifier. Compared to symmetric pure ZGNRs, the insertion of the CAC in these structures strengthens the symmetry restriction and greatly enhances the rectification ratios. Structures with electrodes of pristine ZGNR can work as spin filter and their rectification performance strongly depends on the magnetization configurations. In the FMa configuration, the spin currents do not change when the bias inverses. On the contrary, the spin current flips when the bias inverses in the FMp configuration. When the edge C atoms in one of the two ZGNR electrodes are substituted by B atoms, the magnetization of this electrode is suppressed and symmetry matched wave functions exist only in one energy band below the Fermi energy. The electrons can pass through the structures only when this electrode is negatively biased and the current rectification ratio may reach over $10^6$. In addition, the current is spin polarized under low bias and becomes unpolarized as the bias increases.

**Acknowledgment**

We appreciate Lei Shen for helpful discussion. This work was supported by the National Natural Science Foundation in China (Grant Nos. 11074182 and 91121021) and a Program for graduates Research & Innovation in University of Jiangsu Province (CXZZ13 0797). It is partially supported by the Qing Lan Project Funded by the Priority Academic Program Development of Jiangsu Higher Education Institutions.

**Author contributions**

Y.J.D. carried out the calculation. Y.J.D. and X.F.W. formulated the ideas of the project, data analysis and the manuscript preparation. X.F.W, S.W.Y., and X.M.W. were responsible for the project direction and manuscript finalization.

**Competing financial interests**

The author(s) declare no competing financial interests.